\documentclass
[aps,pra,amsfonts,amssymb,twocolumn,amsmath,preprintnumbers,nofootinbib,floatfix,
showpacs]{revtex4-1}%
\usepackage[dvips]{graphics}
\usepackage{graphicx}
\usepackage{bm}
\usepackage{amsmath}
\usepackage{amsfonts}
\usepackage{amssymb}
\usepackage{xcolor}
\usepackage{subfigure}
\usepackage{hyperref,hypcap}
\usepackage{braket}
\usepackage{commath}%
\providecommand{\U}[1]{\protect\rule{.1in}{.1in}}

\begin{document}

\title{Unified bulk semiclassical theory for intrinsic thermal transport and
magnetization currents}
\author{Cong Xiao}
\author{Qian Niu}
\affiliation{Department of Physics, The University of Texas at Austin, Austin, Texas 78712, USA}

\begin{abstract}
We reveal the unexpected role of the material inhomogeneity in unifying the
formulation of intrinsic thermal and thermoelectric transport as well as
magnetization currents. The smooth inhomogeneity leads to the position
dependent local band dispersion and \textit{phase-space} Berry curvature,
enabling a general and rapid access to transport and magnetization currents
displaying the \textit{momentum-space} Berry curvature physics. Our theory
does not invoke the boundary current, the thermodynamic approach to
magnetization or any mechanical counterpart of statistical forces. By
introducing a fictitious inhomogeneity, it applies to homogeneous samples as
well, promoting the inhomogeneity to be a basic trick in semiclassical
transport theories. Such a trick works regardless of the driving force of
transport, e.g., temperature gradient, in contrast to the trick of fictitious
gravitational field in quantum transport theories. We thus include more
general mechanical driving forces and establish the Mott relation between the
resulting transport thermal and electric currents, whereas this relation for
these two currents was previously only known when an electric field is the
driving force.
\end{abstract}
\maketitle


\section{Introduction}

Momentum-space Berry curvature effects in various nonequilibrium phenomena in
crystalline solids driven by statistical forces, namely the gradients of
temperature and chemical potential ($\boldsymbol{\nabla}T$ and
$\boldsymbol{\nabla}\mu$), have been extensively studied in recent years.
Prominent examples include the anomalous and spin Nernst effects
\cite{Xiao2006,Zhang2008,Jauho2015,Cheng2016} and the thermal Hall effect
\cite{Murakami2011PRB,Murakami2011PRL,Zhang2016,Zhang2010,Qin2012,Murakami2014,Zhang2019,Park2019,Saito2019,Go2019,Lee2015}%
. A basic issue in these phenomena is that the transport current of
experimental interest differs from the local one by the magnetization current,
which cannot be measured in standard transport experiments
\cite{Cooper1997,Xiao2006}. An efficient, intuitive and systematic approach to
subtracting the magnetization current from the local one is thus vital for
understanding the anomalous thermoelectric and thermal transport.

Quantum theories have been formulated to address the aforementioned phenomena.
For this purpose, a fictitious gravitational potential
\cite{Luttinger1964,Streda1977,Qin2011,Murakami2011PRB,Murakami2011PRL,Qin2012,Murakami2014,Lee2015}
or a thermal vector potential \cite{Tatara2015} has to be introduced as the
mechanical counterpart of the temperature gradient. Alternatively the gauge
theory of gravity (Cartan geometry) combined with the Keldysh perturbation
formalism is employed, describing the magnetization of the thermal current
thermodynamically as a torsional response \cite{Shitade2014}. Nevertheless,
such approaches are usually technically complicated and less intuitive
compared to the semiclassical ones.

Semiclassical theories, based on the Berry-curvature-modified semiclassical
equations of motion of the carrier wave-packets, are intuitive and thus found
many applications in the intrinsic anomalous thermoelectric and thermal Hall
effects
\cite{Xiao2006,Zhang2008,Murakami2011PRB,Murakami2011PRL,Zhang2016,Zhang2019,Park2019,Go2019}%
. However, when the thermal current response is considered, a thorough and
systematic understanding of its transport and magnetization parts from the
semiclassical theory is still lacking. On one hand, the existing semiclassical
theory on the basis of bulk considerations \cite{Xiao2006} can only deal with
the thermal current induced by the electric field $\boldsymbol{E}$, but cannot
accommodate that induced by statistical forces. On the other hand, the
semiclassical theory based on the boundary-current picture
\cite{Murakami2011PRB,Murakami2011PRL} cannot include the effect of a uniform
electric field. And how to distinguish the transport and magnetization
currents in this approach is not apparent. Besides, these theories assume no
bulk material inhomogeneity, which may not be true in realistic samples
\cite{Cooper1997}. For instance, in ferromagnets where the anomalous and
thermal Hall effects appear, the magnetization order parameter field, at the
mean-field level, may change slowly in space \cite{Xiao2009}.

Moreover, even in homogeneous samples there are subtleties in the physical
interpretation of existing semiclassical theories
\cite{Xiao2006,Murakami2011PRB,Murakami2011PRL}. Noticeably, in the bulk
semiclassical theory the intrinsic (i.e., determined solely by band
structures) transport electric current induced by ($\boldsymbol{\nabla}\mu$,
$\boldsymbol{\nabla}T$) comes solely from the orbital magnetization current
\cite{Xiao2006}, seeming to conflict with the basic concept that the
magnetization current does not contribute to transport \cite{Cooper1997}. On
the other hand, in the boundary-current theory the obtained current takes the
form of a curl \cite{Murakami2011PRB,Murakami2011PRL}, resembling the
magnetization current \cite{Cooper1997} instead.

In this paper, based on pure bulk considerations we formulate a semiclassical
theory for the thermal and electric current responses in the presence of
($\boldsymbol{E}$, $\boldsymbol{\nabla}\mu$, $\boldsymbol{\nabla}T$) as well
as material inhomogeneity. Assumed to be smooth in the spread of a
wave-packet, the inhomogeneity fits naturally into the semiclassical theory
due to the locality of wave-packet \cite{Sundaram1999}, but this advantage has
not been fully exploited in previous considerations on thermal transport.
Notably, we show that introducing the smooth inhomogeneity serves as a general
and rapid access to transport and magnetization currents displaying the
momentum-space Berry curvature physics. As a result, this theory does not need
to appeal to the boundary-current picture, the thermodynamic and
electrodynamic approaches to various (orbital, particle, energy and thermal)
magnetizations \cite{Xiao2006,Shi2007,Shitade2014} or any mechanical
counterpart of statistical forces.

Our theory also applies conveniently to homogeneous samples by introducing an
fictitious inhomogeneity, thus providing a unified treatment for thermal
transport in both inhomogeneous and homogeneous systems. Such a treatment
avoids aforementioned subtleties in the physical interpretation of the widely
referred semiclassical theories in homogeneous samples
\cite{Xiao2006,Murakami2011PRB,Murakami2011PRL}.

Our finding promotes the (fictitious) smooth inhomogeneity to be a basic and
generic trick in the semiclassical transport theory. Unlike the trick of the
fictitious gravitational field
\cite{Luttinger1964,Streda1977,Qin2011,Murakami2014} which mimics the
temperature gradient in quantum transport theories, the trick of inhomogeneity
is not proposed to mimic any driving force of transport, thus is not limited
to the transport induced by ($\boldsymbol{E}$, $\boldsymbol{\nabla}\mu$,
$\boldsymbol{\nabla}T$). To show this generality, we work out the linear
electric and thermal transport driven by the first-order spatial gradient of
any gauge-invariant static vector field $\boldsymbol{F}$ that perturbs the
single-particle Hamiltonian in the form of $\boldsymbol{\hat{\theta}}%
\cdot\boldsymbol{F}\left(  \boldsymbol{r}\right)  $. Here $\boldsymbol{\hat
{\theta}}$ is the observable coupled to the $\boldsymbol{F}$ field, and is
assumed to be well-defined in the Bloch representation. A particular example
covered is the thermoelectric transport driven by the gradient of an external
Zeeman field. To study transport we need to subtract the orbital and thermal
magnetization currents in the presence of the aforementioned perturbation,
which have not been addressed in previous theories. We establish the Mott
relation between the transport thermal and electric currents induced by the
generic driving forces, whereas, for these two kinds of current, textbooks
\cite{Ziman1972,Ashcroft} and previous studies only proved the Mott relation
when an electric field is the driving force.

The rest of the paper is organized as follows. Section II is devoted to the
preliminaries of the magnetization current and the semiclassical wave-packet
dynamics under nonuniform circumstances. In Sec. III we set forth the theory
for obtaining the magnetization current in the absence of driving forces. The
theory is then extended to involve both the magnetization and transport
currents in the presence of statistical forces and the electric field in Sec.
IV. We show that this approach applies to homogeneous samples as well in Sec.
V. Finally, we include more general driving forces into our theory in Sec. VI,
and conclude the paper in Sec. VII.

\section{Preliminaries}

\subsection{Local current and magnetization current}

The local thermal current density is given by
\begin{equation}
\boldsymbol{j}^{\text{h}}\left(  \boldsymbol{r}\right)  \equiv\boldsymbol{j}%
^{\text{E}}\left(  \boldsymbol{r}\right)  -\mu\left(  \boldsymbol{r}\right)
\boldsymbol{j}^{\text{N}}\left(  \boldsymbol{r}\right)  ,
\end{equation}
where $\boldsymbol{j}^{\text{E}}\left(  \boldsymbol{r}\right)  $ and
$\boldsymbol{j}^{\text{N}}\left(  \boldsymbol{r}\right)  $ are the local
energy current and local particle current densities, respectively. Since
$\boldsymbol{j}^{\text{E}}$ and $\boldsymbol{j}^{\text{N}}$ are conserved
currents, there can be a circulating component that is a curl of some bulk
quantity and cannot be measured in transport experiments \cite{Cooper1997}. If
the quasi-particle number is not conserved, $\boldsymbol{j}^{\text{N}}$ is not
well defined but $\mu=0$, thus $\boldsymbol{j}^{\text{h}}\left(
\boldsymbol{r}\right)  \equiv\boldsymbol{j}^{\text{E}}\left(  \boldsymbol{r}%
\right)  $. For carriers with a conserved charge, say, electrons with charge
$e$, the particle current implies a charge one $\boldsymbol{j}^{\text{e}%
}\left(  \boldsymbol{r}\right)  =e\boldsymbol{j}^{\text{N}}\left(
\boldsymbol{r}\right)  $, whose circulating component is the orbital
magnetization current $\boldsymbol{j}^{\text{e,mag}}=\boldsymbol{\nabla}%
\times\boldsymbol{M}^{\text{e}}$ according to electromagnetism, with
$\boldsymbol{M}^{\text{e}}$ being the orbital magnetization. Given this
convention, the circulating energy and particle currents can be termed as
magnetization currents $\boldsymbol{j}^{\text{E,mag}}=\boldsymbol{\nabla
}\times\boldsymbol{M}^{\text{E}}$ and $\boldsymbol{j}^{\text{N,mag}%
}=\boldsymbol{\nabla}\times\boldsymbol{M}^{\text{N}}$, with $\boldsymbol{M}%
^{\text{E}}$\ and $\boldsymbol{M}^{\text{N}}$ being respectively the energy
magnetization and particle magnetization, albeit the thermodynamic definitions
of these magnetizations\ are not apparent. The local current density is thus
composed of the transport and magnetization parts: $\boldsymbol{j}%
^{\text{E(N)}}\left(  \boldsymbol{r}\right)  =\boldsymbol{j}^{\text{E(N),tr}%
}\left(  \boldsymbol{r}\right)  +\boldsymbol{j}^{\text{E(N),mag}}\left(
\boldsymbol{r}\right)  $, and then%
\begin{equation}
\boldsymbol{j}^{\text{h}}\left(  \boldsymbol{r}\right)  =\boldsymbol{j}%
^{\text{h,tr}}\left(  \boldsymbol{r}\right)  +\boldsymbol{j}^{\text{h,mag}%
}\left(  \boldsymbol{r}\right)  . \label{local heat current}%
\end{equation}
Here the thermal magnetization current density $\boldsymbol{j}^{\text{h,mag}%
}\equiv\boldsymbol{j}^{\text{E,mag}}-\mu\boldsymbol{j}^{\text{N,mag}}$ is
given by
\begin{equation}
\boldsymbol{j}^{\text{h,mag}}=\boldsymbol{\nabla}\times(\boldsymbol{M}%
^{\text{E}}-\mu\boldsymbol{M}^{\text{N}})+\boldsymbol{\nabla}\mu
\times\boldsymbol{M}^{\text{N}}, \label{heat-M current}%
\end{equation}
which is not simply a curl of some thermal magnetization in the presence of
statistical inhomogeneity.

Before proceeding, we outline the strategy of our theory for acquiring the
magnetization and transport currents. The semiclassical wave-packet theory
introduced shortly allows for acquiring the local current densities, which
reduce to the magnetization ones in the absence of any statistical and
mechanical driving forces. Hence we will first obtain the magnetization
current in this case, based on which we can go further to identify the
magnetization current in the presence of the statistical and mechanical
driving forces. The transport current are thus reached by subtracting the
magnetization current from the local one.

\subsection{Semiclassical description in nonuniform bulk}

In the semiclassical theory \cite{Xiao2010,Ashcroft}, a Bloch electron is
physically identified as a wave-packet $|\Phi\left(  \boldsymbol{q}%
_{c},\boldsymbol{r}_{c},t\right)  \rangle$ that is constructed from the Bloch
states in a particular nondegenerate band (throughout this paper the band
index $n$ is usually omitted for simplicity, unless otherwise noted) and is
localized around a central position $\boldsymbol{r}_{c}$ and a mean crystal
momentum $\boldsymbol{q}_{c}$. Assuming all the inhomogeneous fields are
static and vary slowly on the spread of the wave-packet, a local Hamiltonian
description for an electron wave-packet emerges, from which the semiclassical
equations of motion are derived \cite{Sundaram1999} (we set $\hbar=1$):
\begin{align}
\dot{\boldsymbol{r}}_{c}=  &  \partial_{\boldsymbol{q}_{c}}\left[
\varepsilon+e\phi\left(  \boldsymbol{r}_{c}\right)  \right]  -\Omega
_{\boldsymbol{q}_{c}\boldsymbol{r}_{c}}\cdot\dot{\boldsymbol{r}}_{c}%
-\Omega_{\boldsymbol{q}_{c}\boldsymbol{q}_{c}}\cdot\dot{\boldsymbol{q}}%
_{c},\nonumber\\
\dot{\boldsymbol{q}}_{c}=  &  -\partial_{\boldsymbol{r}_{c}}\left[
\varepsilon+e\phi\left(  \boldsymbol{r}_{c}\right)  \right]  +\Omega
_{\boldsymbol{r}_{c}\boldsymbol{r}_{c}}\cdot\dot{\boldsymbol{r}}_{c}%
+\Omega_{\boldsymbol{r}_{c}\boldsymbol{q}_{c}}\cdot\dot{\boldsymbol{q}}_{c}.
\label{EOM}%
\end{align}
Here $\left(  \Omega_{\boldsymbol{\lambda\lambda}}\right)  _{ij}%
=2\operatorname{Im}\langle\partial_{\lambda_{j}}u|\partial_{\lambda_{i}%
}u\rangle$ are the Berry curvatures derived from the periodic part
$|u(\boldsymbol{q}_{c},\boldsymbol{r}_{c})\rangle$ of the local Bloch wave
function, where $i$ and $j$ are Cartesian indices, and $\boldsymbol{\lambda
}=\boldsymbol{r}_{c}$, $\boldsymbol{q}_{c}$.

To study the magnetization current which only manifests itself in the presence
of inhomogeneity, we introduce a slowly-varying nonuniform field
$\boldsymbol{w}\left(  \boldsymbol{r}\right)  $ to represent the material
inhomogeneity. Hence the spatial derivative $\partial_{\boldsymbol{r}_{c}}$ in
the equations of motion acts through $\boldsymbol{w}\left(  \boldsymbol{r}%
_{c}\right)  $. This is a route to manifest the magnetization current in bulk
even in the global equilibrium without position dependent temperature or
chemical potential, outside of the scope of previous semiclassical theories
\cite{Xiao2006,Murakami2011PRB,Murakami2011PRL,Zhang2016}. The $\boldsymbol{w}%
$ field can exist indeed in the system, representing the realistic material
inhomogeneity, or just be an auxiliary tool in a homogeneous sample. In the
former case, a specific realization is the slowly-varying spin texture in
ferromagnets with an inhomogeneous equilibrium magnetization order parameter
field \cite{Xiao2009}. In the latter case the fictitious $\boldsymbol{w}$
field will be dropped after identifying the magnetization current. In both
cases the specific content of $\boldsymbol{w}$ field is not needed in our
theory. It can be a scalar, vector or tensor field. When it is a scalar field,
the local band energy is shifted but the local Bloch wave-function is not
affected, hence $\Omega_{\boldsymbol{q}_{c}\boldsymbol{r}_{c}}=\Omega
_{\boldsymbol{r}_{c}\boldsymbol{r}_{c}}=0$, simplifying the analysis. However,
as mentioned above, the material inhomogeneity may appear as a nonuniform
vector field. Therefore, in what follows we assume $\boldsymbol{w}$ field is
not a scalar one for the sake of generality.

In the equations of motion, $\varepsilon+e\phi\left(  \boldsymbol{r}%
_{c}\right)  $ is the wave-packet energy, where $\varepsilon\equiv
\varepsilon(\boldsymbol{q}_{c},\boldsymbol{r}_{c})=\varepsilon(\boldsymbol{q}%
_{c},\boldsymbol{w}(\boldsymbol{r}_{c}))$ is the local band dispersion without
the coupling to driving fields that do not equilibrate with the system, such
as the electrostatic potential $e\phi\left(  \boldsymbol{r}_{c}\right)  $. The
dipole moment correction from the gradient of $\boldsymbol{w}\left(
\boldsymbol{r}_{c}\right)  $ to the wave-packet energy \cite{Sundaram1999} is
not essential for the present topic, as can be easily verified.

Within the validity of the uncertainty principle, the phase-space occupation
function $f\left(  \boldsymbol{q}_{c},\boldsymbol{r}_{c}\right)  $ of a grand
canonical ensemble of dynamically independent semiclassical Bloch electrons
can be defined, and the phase-space measure $D\left(  \boldsymbol{q}%
_{c},\boldsymbol{r}_{c}\right)  $ has to be introduced. Because of the
non-canonical structure of the equations of motion (\ref{EOM}) shaped by the
Berry curvatures \cite{Xiao2005}, $D\neq1$ and reads%
\begin{equation}
D\left(  \boldsymbol{q}_{c},\boldsymbol{r}_{c}\right)  =1+\left(
\Omega_{\boldsymbol{q}_{c}\boldsymbol{r}_{c}}\right)  _{ii} \label{DOS}%
\end{equation}
up to the first order of spatial gradient. Summation over repeated Cartesian
indices is implied henceforth. The number of states within a small phase-space
volume is hence given by $Dfd\boldsymbol{r}_{c}d\boldsymbol{q}_{c}/(2\pi)^{d}%
$, with $d$ as the spatial dimensionality. In this paper we do not consider
the off equilibrium distribution function, as the pertinent transport
contributions can be described by the Boltzmann equation \cite{Ashcroft}.
Therefore, in the following $f$ is just the equilibrium occupation function,
which is given by the Fermi-Dirac distribution function $f\left(
\varepsilon\right)  \equiv f_{DF}\left(  \varepsilon(\boldsymbol{q}%
_{c},\boldsymbol{r}_{c})\right)  $ in the case of Bloch electrons.

\section{Magnetization current at global equilibrium}

First we look at the case of global equilibrium without any statistical or
mechanical force driving nonequilibrium states. As is pointed out above, in
the presence of material inhomogeneity represented by a nonuniform
$\boldsymbol{w}$ field, $\boldsymbol{j}^{\text{E}}=\boldsymbol{j}%
^{\text{E,mag}}\neq0$ and $\boldsymbol{j}^{\text{N}}=\boldsymbol{j}%
^{\text{N,mag}}\neq0$ in the bulk. Thus
\begin{equation}
\boldsymbol{j}^{\text{h}}=\boldsymbol{j}^{\text{h,mag}}=\boldsymbol{\nabla
}\times\boldsymbol{M}^{\text{h}}, \label{heat-M eq}%
\end{equation}
where $\boldsymbol{M}^{\text{h}}=\boldsymbol{M}^{\text{E}}-\mu\boldsymbol{M}%
^{\text{N}}$ is the thermal magnetization in the absence of electric fields,
and $\mu$ is a constant.

In the semiclassical theory the local energy current density reads%
\begin{equation}
j_{i}^{\text{E}}\left(  \boldsymbol{r}\right)  \equiv\int\left[
d\boldsymbol{q}_{c}\right]  d\boldsymbol{r}_{c}Df\left(  \varepsilon\right)
\varepsilon\langle\Phi|\hat{v}_{i}\delta\left(  \hat{\boldsymbol{r}%
}-\boldsymbol{r}\right)  |\Phi\rangle, \label{local-jE}%
\end{equation}
and the local particle current density is%
\begin{equation}
j_{i}^{\text{N}}\left(  \boldsymbol{r}\right)  \equiv\int\left[
d\boldsymbol{q}_{c}\right]  d\boldsymbol{r}_{c}Df\left(  \varepsilon\right)
\langle\Phi|\hat{v}_{i}\delta\left(  \hat{\boldsymbol{r}}-\boldsymbol{r}%
\right)  |\Phi\rangle. \label{local-jN}%
\end{equation}
Here $\left[  d\boldsymbol{q}_{c}\right]  $ is shorthand for $\sum
_{n}d\boldsymbol{q}_{c}/\left(  2\pi\right)  ^{d}$, and $\hat{\boldsymbol{v}}$
and $\hat{\boldsymbol{r}}$ are respectively the velocity and position
operators. Expanding the $\delta$\ function to first order of $\hat
{\boldsymbol{r}}-\boldsymbol{r}_{c}$ yields
\begin{align}
j_{i}^{\text{h}}\left(  \boldsymbol{r}\right)  =  &  \int Df\left(
\varepsilon\right)  \left(  \varepsilon-\mu\right)  \langle\Phi|\hat{v}%
_{i}|\Phi\rangle|_{\boldsymbol{r}_{c}=\boldsymbol{r}}\label{local-jh}\\
&  -\partial_{r_{j}}\int f\left(  \varepsilon\right)  \left(  \varepsilon
-\mu\right)  \langle\Phi|\hat{v}_{i}\left(  \hat{r}_{j}-r_{j}\right)
|\Phi\rangle|_{\boldsymbol{r}_{c}=\boldsymbol{r}}.\nonumber
\end{align}
Henceforth we omit the center position label $c$, unless otherwise noted, and
the notation $\int$ without integral variable is shorthand for $\int\left[
d\boldsymbol{q}_{c}\right]  $. We are limited to the first order of spatial
gradients, thus it is sufficient to set $D=1$\ in the second term in the above expansion.

To proceed, we introduce two functions $g\left(  \varepsilon,\mu,T\right)  $
and $h\left(  \varepsilon,\mu,T\right)  $ which satisfy%
\begin{equation}
\frac{\partial g}{\partial\varepsilon}=f\left(  \varepsilon\right)  ,\text{
\ }\frac{\partial h}{\partial\varepsilon}=f\left(  \varepsilon\right)  \left(
\varepsilon-\mu\right)  .
\end{equation}
In fact $g\left(  \varepsilon\right)  =-k_{B}T\ln[1+e^{-(\varepsilon
-\mu)/k_{B}T}]$ is the grand potential density contributed by a particular
state, whereas $h=-\int_{\varepsilon}^{\infty}d\eta f\left(  \eta\right)
\left(  \eta-\mu\right)  $. Then the local thermal (particle) current density
reads \cite{Xiao2006,Zhang2016}%
\begin{equation}
\boldsymbol{j}^{\text{h(N)}}=\int D\frac{\partial\Lambda}{\partial\varepsilon
}\dot{\boldsymbol{r}}+\boldsymbol{\nabla}\times\int\frac{\partial\Lambda
}{\partial\varepsilon}\boldsymbol{m}^{\text{N}}\text{ \ for \ }\Lambda
=h\left(  g\right)  . \label{local density}%
\end{equation}
The first and second terms come from the motion of the wave-packet center with
velocity $\dot{\boldsymbol{r}}=\langle\Phi|\hat{\boldsymbol{v}}|\Phi\rangle$
and the wave-packet self-rotation, respectively. $\boldsymbol{m}^{\text{N}}$
is the particle magnetic moment, which is the vector form of the antisymmetric
tensor $m_{ji}^{\text{N}}=\langle\Phi|\hat{v}_{i}\left(  \hat{r}_{j}%
-r_{j}\right)  |\Phi\rangle$ (symmetrization between operators is implied)
\cite{Xiao2010}. For $\boldsymbol{j}^{\text{h}}$ the second term reads
alternatively $\boldsymbol{\nabla}\times\int f\boldsymbol{m}^{\text{h}}$, with
$\boldsymbol{m}^{\text{h}}=\left(  \varepsilon-\mu\right)  \boldsymbol{m}%
^{\text{N}}$ being the so-called thermal magnetic moment \cite{Zhang2016}.

Above expressions for the local current are well known. However, now we are in
a position to acquire the thermal magnetization current from a pure bulk
consideration, outside of the scope of the previous bulk semiclassical theory
\cite{Xiao2006}. In fact, in the latter the local currents vanish
$\boldsymbol{j}^{\text{h(N)}}=0$ in the present case without statistical and
mechanical driving forces.

Plugging the equations of motion (\ref{EOM}) in the first order of spatial
gradients and the phase-space measure (\ref{DOS}) into the first term of
(\ref{local density}) and making use of $\boldsymbol{\nabla}\Lambda
=\partial_{\varepsilon}\Lambda\boldsymbol{\nabla}\varepsilon$, we arrive at%
\begin{equation}
\int D\frac{\partial\Lambda}{\partial\varepsilon}\dot{\boldsymbol{r}%
}=\boldsymbol{\nabla}\times\int\Lambda\left(  \varepsilon\right)
\boldsymbol{\Omega}_{\boldsymbol{q}},\text{ (}\Lambda=g,h\text{),}
\label{M edge}%
\end{equation}
with $\left(  \Omega_{\boldsymbol{q}}\right)  _{k}=\frac{1}{2}\epsilon
_{ijk}\left(  \Omega_{\boldsymbol{q}\boldsymbol{q}}\right)  _{ij}$ being the
vector form of the momentum-space Berry curvature.\ This is one of the pivotal
results of our approach. Since $\dot{\boldsymbol{r}}$ represents the motion of
the wave-packet center, it may not be very apparent to envision that $\int
D\frac{\partial\Lambda}{\partial\varepsilon}\dot{\boldsymbol{r}}$ can be
totally a part of the magnetization current. According to the above two
equations we get%
\begin{equation}
\boldsymbol{j}^{\text{h(N)}}=\boldsymbol{\nabla}\times\int(\frac
{\partial\Lambda}{\partial\varepsilon}\boldsymbol{m}^{\text{N}}+\Lambda
\boldsymbol{\Omega}_{\boldsymbol{q}})\text{ \ for }\Lambda=h\left(  g\right)
\text{.}%
\end{equation}
Since they are just the magnetization currents, we identify $\boldsymbol{M}%
^{\text{h(N)}}=\int(\frac{\partial\Lambda}{\partial\varepsilon}\boldsymbol{m}%
^{\text{N}}+\Lambda\boldsymbol{\Omega}_{\boldsymbol{q}})$ for $\Lambda
=h\left(  g\right)  $ as the thermal (particle) magnetization up to a
gradient. Our result for $\boldsymbol{M}^{\text{N}}$ is consistent with the
thermodynamic one \cite{Xiao2006,Shi2007}, and the obtained $\boldsymbol{M}%
^{\text{h}}$ coincides with that obtained using the gauge theory of gravity
\cite{Shitade2014}, where the thermal magnetization is thermodynamically
defined as the derivative of the grand potential with respect to the torsional
magnetic field, further confirming the validity of our theory.

\section{Transport and magnetization currents in the presence of driving
forces}

In the presence of statistical forces, the local current density consists of
both the magnetization and transport parts. Manipulations similar to Eq.
(\ref{local-jh}) lead to the local thermal current density%
\begin{equation}
\boldsymbol{j}^{\text{h}}=\int D\frac{\partial h}{\partial\varepsilon}%
\dot{\boldsymbol{r}}+\boldsymbol{\nabla}\times\int\frac{\partial h}%
{\partial\varepsilon}\boldsymbol{m}^{\text{N}}+\boldsymbol{\nabla}\mu
\times\int f\boldsymbol{m}^{\text{N}}.
\end{equation}
Taking some technical steps similar to those of Eq. (\ref{M edge}) and
noticing that $\boldsymbol{\nabla}h=\frac{\partial h}{\partial\varepsilon
}\boldsymbol{\nabla}\varepsilon+\frac{\partial h}{\partial T}%
\boldsymbol{\nabla}T+\frac{\partial h}{\partial\mu}\boldsymbol{\nabla}\mu$ in
the present case, we get%
\begin{equation}
\int D\frac{\partial h}{\partial\varepsilon}\dot{\boldsymbol{r}}%
=-(\boldsymbol{\nabla}\mu\times\frac{\partial}{\partial\mu}+\boldsymbol{\nabla
}T\times\frac{\partial}{\partial T})\int h\boldsymbol{\Omega}_{\boldsymbol{q}%
}+\boldsymbol{\nabla}\times\int h\boldsymbol{\Omega}_{\boldsymbol{q}},
\end{equation}
which is another pivotal result of our approach. It is of interest because, as
long as the material inhomogeneity is absent one would only get $\int
D\frac{\partial h}{\partial\varepsilon}\dot{\boldsymbol{r}}=0$, even in the
presence of statistical forces. However, introducing the material
inhomogeneity enables us to show that the motion of the wave-packet center in
bulk can contribute to both the transport and magnetization currents.

Then the local thermal current density is given by%
\begin{align}
\boldsymbol{j}^{\text{h}}  &  =\boldsymbol{\nabla}\mu\times\int
\boldsymbol{\Omega}_{\boldsymbol{q}}s\left(  \varepsilon\right)
T-\boldsymbol{\nabla}T\times\int\frac{\partial h}{\partial T}%
\boldsymbol{\Omega}_{\boldsymbol{q}}\nonumber\\
&  +\boldsymbol{\nabla}\times\boldsymbol{M}^{\text{h}}+\boldsymbol{\nabla}%
\mu\times\boldsymbol{M}^{\text{N}}. \label{jh loc-eq}%
\end{align}
Here $\boldsymbol{M}^{\text{h}}$ and $\boldsymbol{M}^{\text{N}}$\ take the
same expressions as those at global equilibrium, with the only difference that
$\mu$ and $T$ are now position dependent at local equilibrium. The second line
of Eq. (\ref{jh loc-eq}) is thus the zero-electric-field thermal magnetization
current (\ref{heat-M current}). The first line of Eq. (\ref{jh loc-eq}) is
then identified as the transport thermal current according to Eq.
(\ref{local heat current}). Here $s\left(  \varepsilon\right)  =\left[
\left(  \varepsilon-\mu\right)  f\left(  \varepsilon\right)  -g\left(
\varepsilon\right)  \right]  /T$ is the entropy density for a particular state.

In the presence of an electric field as well as statistical forces, the local
energy current density is still given by Eq. (\ref{local-jE}), but with the
difference that the transported carrier energy changes to be $\varepsilon
+e\phi\left(  \boldsymbol{r}_{c}\right)  $, as shown in the dynamic equations
(\ref{EOM}). Note that the electric field does not equilibrate with the
electron system thus the equilibrium phase-space distribution function remains
as $f\left(  \varepsilon\right)  $. Manipulations similar to what have been
done yield the local thermal current density as%
\begin{align}
\boldsymbol{j}^{\text{h}}  &  =\left(  e\boldsymbol{E}-\boldsymbol{\nabla}%
\mu\right)  \times T\int\frac{\partial g}{\partial T}\boldsymbol{\Omega
}_{\boldsymbol{q}}-\boldsymbol{\nabla}T\times\int\frac{\partial h}{\partial
T}\boldsymbol{\Omega}_{\boldsymbol{q}}\nonumber\\
&  +\boldsymbol{\nabla}\times\left[  \boldsymbol{M}^{\text{h}}+e\phi
\boldsymbol{M}^{\text{N}}\right]  +\boldsymbol{\nabla}\mu\times\boldsymbol{M}%
^{\text{N}}. \label{jh-local}%
\end{align}
Here $\boldsymbol{M}^{\text{h}}$ is the zero-electric-field thermal
magnetization, and we have used the relation $s=-\partial g/\partial T$.

The presence of the $e\phi\boldsymbol{M}^{\text{N}}$ term just reflects the
result in the quantum mechanical linear response theories that the energy
magnetization becomes $\boldsymbol{M}^{\text{E}}+e\phi\boldsymbol{M}%
^{\text{N}}$ in the presence of the electrostatic potential \cite{Cooper1997}.
Here $\boldsymbol{M}^{\text{E}}$ stands for the zero-electric-field energy
magnetization. The general form of the bulk thermal magnetization current
given by the second line of (\ref{jh-local}) in the presence of not only
($\phi$, $\boldsymbol{\nabla}\mu$, $\boldsymbol{\nabla}T$) but also the
material inhomogeneity is consistent with the full quantum theory (Eqs. (76)
and (77) in Ref. \cite{Cooper1997}). This achievement has not been reached in
previous semiclassical theories. The potential field $\phi$ is introduced to
produce the electric field $\boldsymbol{E}=-\boldsymbol{\nabla}\phi$, and
after achieving this one can always set $\phi\left(  \boldsymbol{r}%
_{c}\right)  =0$. Therefore the thermal magnetization current can be expressed
as
\begin{equation}
\boldsymbol{j}^{\text{h,mag}}=\boldsymbol{\nabla}\times\boldsymbol{M}%
^{\text{h}}-\left(  e\boldsymbol{E}-\boldsymbol{\nabla}\mu\right)
\times\boldsymbol{M}^{\text{N}}. \label{jh-magnetization}%
\end{equation}
In the absence of the material inhomogeneity ($\boldsymbol{w}$ field) and
statistical forces, $\boldsymbol{M}^{\text{h}}$ is a constant in bulk. Then
$\boldsymbol{j}^{\text{h,mag}}=-\boldsymbol{E}\times\boldsymbol{M}^{\text{e}}%
$, reducing to the result in the previous bulk semiclassical theory
\cite{Xiao2006}. There this magnetization current is obtained on the basis of
a physical argument of the material dependent part of the Poynting vector,
while here such an argument is not needed.

The transport thermal current, according to Eq. (\ref{local heat current}), is
just given by the first line of Eq. (\ref{jh-local}). Noticing that
$\boldsymbol{M}^{\text{e,B}}=e\int g\boldsymbol{\Omega}_{\boldsymbol{q}}$ and
$\boldsymbol{M}^{\text{h,B}}=\int h\boldsymbol{\Omega}_{\boldsymbol{q}}$ are
respectively the Berry-curvature parts of the orbital magnetization
($\boldsymbol{M}^{\text{e}}=e\boldsymbol{M}^{\text{N}}$) and thermal
magnetization, we have
\begin{equation}
\boldsymbol{j}^{\text{h,tr}}=(\boldsymbol{E}-\frac{1}{e}\boldsymbol{\nabla}%
\mu)\times T\frac{\partial\boldsymbol{M}^{\text{e,B}}}{\partial T}%
-\boldsymbol{\nabla}T\times\frac{\partial\boldsymbol{M}^{\text{h,B}}}{\partial
T}. \label{jh-transport}%
\end{equation}
Similarly, the transport electric current reads%
\begin{equation}
\boldsymbol{j}^{\text{e,tr}}=\left(  e\boldsymbol{E}-\boldsymbol{\nabla}%
\mu\right)  \times\frac{\partial\boldsymbol{M}^{\text{e,B}}}{\partial\mu
}-\boldsymbol{\nabla}T\times\frac{\partial\boldsymbol{M}^{\text{e,B}}%
}{\partial T}, \label{je-transport}%
\end{equation}
thus one immediately verifies the Einstein relation, the Onsager relation and
the Wiedemann-Franz law, in the presence of material inhomogeneity. In
insulators $\boldsymbol{M}^{\text{e,B}}$ ($\boldsymbol{M}^{\text{h,B}}$) can
be replaced by $\boldsymbol{M}^{\text{e}}$ ($\boldsymbol{M}^{\text{h}}$) in
Eqs. (\ref{jh-transport}) and (\ref{je-transport}), recovering the Streda
formulas \cite{Streda1983,Nagaosa2012,Zhang2016} which link the transport
coefficients to the derivatives of magnetizations with respect to $\mu$ or $T$.

\section{Application to homogeneous samples}

The obtained transport and magnetization currents [Eqs.
(\ref{jh-magnetization}) to (\ref{je-transport})] are also valid in the
absence of material inhomogeneity. In fact, in homogeneous samples, which all
previous semiclassical thermoelectric and thermal transport theories
\cite{Murakami2011PRB,Murakami2011PRL,Xiao2006,Zhang2016,Cheng2016,Zhang2019}
are designed for, one can introduce a fictitious nonuniform $\boldsymbol{w}$
field that is removed at the last of the calculation to reach the above
results. Our approach thus unifies the treatment and understanding of the
intrinsic thermal transport in both homogeneous and inhomogeneous samples.

In the following we make a comparison of semiclassical thermoelectric
transport theories in the present and previous papers. In the previous bulk
semiclassical thermoelectric transport theory \cite{Xiao2006}, the orbital
magnetization is acquired separately by thermodynamics as the derivative of
the grand potential density with respect to magnetic field, and then the
magnetization current is obtained. This route poses the basic requirement for
a thermodynamic definition of the thermal magnetization, which is not evident
in the familiar context of condensed matter physics. Therefore, the thermal
Hall transport was not touched in Ref. \cite{Xiao2006}. The particle
magnetization current of neutral quasi-particles that do not couple to the
magnetic field by the Lorentz force \cite{Cheng2016} suffers from the same
situation. In contrast, in the present transport approach, the thermal
magnetization and particle magnetization currents emerge naturally.

With the identified thermal magnetization current (\ref{jh-magnetization}),
the result of the boundary-current approach
\cite{Murakami2011PRB,Murakami2011PRL}, which applies to the case with merely
statistical forces, can be derived from the present bulk theory. In this
situation the $\boldsymbol{w}$\ field is absent, thus $\boldsymbol{\nabla
=\nabla}\mu\frac{\partial}{\partial\mu}+\boldsymbol{\nabla}T\frac{\partial
}{\partial T}$, and all the spatial gradients in the equations of motion
vanish, implying $\dot{\boldsymbol{r}}=\partial_{\boldsymbol{q}}\varepsilon$
and $D=1$. The local thermal current density then reduces to $\boldsymbol{j}%
^{\text{h}}=\boldsymbol{\nabla}\times\int f\boldsymbol{m}^{\text{h}%
}+\boldsymbol{\nabla}\mu\times\int f\boldsymbol{m}^{\text{N}}$. Subtracting
the magnetization current (\ref{jh-magnetization}) yields
\[
\boldsymbol{j}^{\text{h,tr}}=-\boldsymbol{\nabla}\times\boldsymbol{M}%
^{\text{E,B}}+\mu\boldsymbol{\nabla}\times\boldsymbol{M}^{\text{N,B}},
\]
with $\boldsymbol{M}^{\text{E,B}}=$ $\int\left(  h+\mu g\right)
\boldsymbol{\Omega}_{\boldsymbol{q}}$ the Berry-curvature part of the energy
magnetization. Concurrently, we prove in the same way $\boldsymbol{j}%
^{\text{E,tr}}=-\boldsymbol{\nabla}\times\boldsymbol{M}^{\text{E,B}}$
and\ $\boldsymbol{j}^{\text{N,tr}}=-\boldsymbol{\nabla}\times\boldsymbol{M}%
^{\text{N,B}}$. These are just the pivotal results of the boundary-current
approach (Eqs. (14) and (15) in Ref. \cite{Murakami2011PRB}). However, the
transport-current nature of $\boldsymbol{j}^{\text{E,tr}}$
and\ $\boldsymbol{j}^{\text{N,tr}}$ is not evident in the boundary-current
approach, since they take the form of a curl, resembling the magnetization
current instead. A straightforward way to clarify this is to introduce a
fictitious material inhomogeneity (i.e., a nonuniform $\boldsymbol{w}$\ field)
and apply our theory, which shows that the intrinsic transport electric
current, for instance, is not essentially a total spatial-derivative, contrary
to the orbital magnetization current, but is always given by Eq.
(\ref{je-transport}).

In Table \ref{correspondence}\ we compare the capabilities of our theory and
previous semiclassical transport theories in homogeneous samples. The driving
forces are not limited to ($\boldsymbol{E}$, $\boldsymbol{\nabla}\mu$,
$\boldsymbol{\nabla}T$), but include also the spatial gradient of a vector
mechanical field $\boldsymbol{F}$, which is detailed in the next section.

\begin{table*}[t]
\caption{The capabilities of our theory and previous representative
semiclassical transport theories of Bloch electrons in homogeneous samples.
$\boldsymbol{j}^{\text{e,tr}}$ and $\boldsymbol{j}^{\text{h,tr}}$ are the
electric and thermal transport currents, respectively. The driving forces
include not only ($\boldsymbol{E}$, $\boldsymbol{\nabla}\mu$,
$\boldsymbol{\nabla}T$), but also the spatial gradient of a vector mechanical
field $\boldsymbol{F}$, which is detailed in the Sec. VI.}%
\label{correspondence}
\centering%
\begin{tabular}
[c]{|c|c|c|c|c|c|}\hline
\ \ semiclassical theories \ \  & \ \ $\boldsymbol{E}\rightarrow
\boldsymbol{j}^{\text{h,tr}}$ \ \  & \ \ $\boldsymbol{\nabla}\mu
,\boldsymbol{\nabla}T\rightarrow\boldsymbol{j}^{\text{e,tr}}$ \ \  &
\ \ $\boldsymbol{\nabla}\mu,\boldsymbol{\nabla}T\rightarrow\boldsymbol{j}%
^{\text{h,tr}}$ \ \  & \ \ $\boldsymbol{\nabla}\boldsymbol{F}\rightarrow
\boldsymbol{j}^{\text{e,tr}}$ \ \  & \ \ $\boldsymbol{\nabla}\boldsymbol{F}%
\rightarrow\boldsymbol{j}^{\text{h,tr}}$\\\hline
previous bulk theory \cite{Xiao2006} & $\checkmark$ & $\checkmark$ &  &
$\checkmark$ & \\\hline
\ \ boundary-current theory \cite{Murakami2011PRB,Murakami2011PRL} \ \  &  &
$\checkmark$ & $\checkmark$ &  & \\\hline
present bulk theory & $\checkmark$ & $\checkmark$ & $\checkmark$ &
$\checkmark$ & $\checkmark$\\\hline
\end{tabular}
\end{table*}

\section{More general mechanical driving forces}

Above our theory has promoted the material inhomogeneity to be a basic trick
in the unified semiclassical theory of the intrinsic linear electric and
thermal transport induced by the driving forces ($\boldsymbol{E}$,
$\boldsymbol{\nabla}\mu$, $\boldsymbol{\nabla}T$). Recall that in the quantum
thermal and thermoelectric transport theories it is the fictitious
gravitational field
\cite{Luttinger1964,Streda1977,Qin2011,Murakami2011PRB,Murakami2011PRL} which
serves as a vital trick to mimic the temperature gradient. However, unlike the
gravitational field, the trick of (fictitious) inhomogeneity has nothing to do
with the driving force of transport, thus is not limited to the transport
induced by ($\boldsymbol{E}$, $\boldsymbol{\nabla}\mu$, $\boldsymbol{\nabla}T$).

In this section we generalize our theory to the linear electric and thermal
transport driven by the spatial gradient of any gauge-invariant vector field
$\boldsymbol{F}$ that perturbs the single-particle Hamiltonian in the form of
$\boldsymbol{\hat{\theta}}\cdot\boldsymbol{F}\left(  \boldsymbol{r}\right)  $.
The genuine or fictitious material inhomogeneity is still represented by the
nonuniform $\boldsymbol{w}$ field. Here $\boldsymbol{\hat{\theta}}$ is the
physical observable operator coupled to the $\boldsymbol{F}$ field, and is
assumed to be well-defined in the Bloch representation. A specific example of
$\boldsymbol{\hat{\theta}}\cdot\boldsymbol{F}$\ is the Zeeman coupling
$\boldsymbol{\hat{S}}\cdot\boldsymbol{Z}\left(  \boldsymbol{r}\right)  $,
where $\boldsymbol{\hat{S}}$ is the carrier spin operator,
and$\ \boldsymbol{Z}\left(  \boldsymbol{r}\right)  $ is an external Zeeman
field. This field does not equilibrate with the electron system and its
spatial gradient drives transport.\ In fact, both $\hat{\boldsymbol{\theta}}$
and $\boldsymbol{F}$ can be tensors in general, and the product denotes the
contraction between them. Thus $\boldsymbol{\hat{\theta}}\cdot\boldsymbol{F}$
also includes the specific case where $\boldsymbol{\hat{\theta}}=e$ is the
carrier charge and $\boldsymbol{F}=\phi$ is the electrostatic potential and
hence the transport is driven by the electric field. Only the spatial
gradients of $\boldsymbol{F}\left(  \boldsymbol{r}\right)  $ matter and are
assumed to be uniform, and one can always choose $\boldsymbol{F}\left(
\boldsymbol{r}_{c}\right)  =0$ at the last of the linear response calculation.

The local electric current density is given by $\boldsymbol{j}^{\text{e}%
}=e\boldsymbol{j}^{\text{N}}$ and Eq. (\ref{local density}). The perturbed
wave-packet energy needed in the calculation reads $\varepsilon
+\boldsymbol{\theta}\cdot\boldsymbol{F}\left(  \boldsymbol{r}_{c}\right)
+\delta\varepsilon$, where $\boldsymbol{\theta}=\langle\Phi|\boldsymbol{\hat
{\theta}}|\Phi\rangle$, and\ ($\partial_{i}\equiv\partial_{r_{i}}$)
\begin{equation}
\delta\varepsilon=d_{ij}^{\theta}\partial_{i}F_{j}%
\end{equation}
is the energy correction induced by the dipole moment of operator
$\boldsymbol{\hat{\theta}}$ on a finite-size wave-packet \cite{Xiao2010}. This
dipole moment takes the explicit form of
\begin{equation}
d_{ij}^{\theta}=\operatorname{Im}\sum_{n^{\prime}\neq n}\frac{\langle
u_{n}|\hat{v}_{i}|u_{n^{\prime}}\rangle\langle u_{n^{\prime}}|\hat{\theta}%
_{j}|u_{n}\rangle}{\varepsilon_{n}-\varepsilon_{n^{\prime}}}, \label{dipole}%
\end{equation}
where $n$ is the index of the band we are considering, and $n^{\prime}$
denotes other bands in the system. In the case where $\boldsymbol{\hat{\theta
}}=e$ is the carrier charge and $\boldsymbol{F}=\phi$ is the electrostatic
potential, we have $\delta\varepsilon=0$ since the electric dipole moment of a
wave-packet is zero, and the perturbed wave-packet energy reduces to that used
in Eq. (\ref{EOM}). In the present more general case, up to the first order of
spatial gradients we obtain
\begin{align*}
\int Df\left(  \varepsilon\right)  \dot{r}_{i}  &  =\int f\left(
\varepsilon\right)  [\partial_{q_{i}}\left(  \varepsilon+\delta\varepsilon
\right)  +\Omega_{q_{j}r_{j}}\partial_{q_{i}}\varepsilon\\
&  +\Omega_{q_{i}q_{j}}(\partial_{j}\varepsilon+\theta_{l}^{0}\partial
_{j}F_{l})-\Omega_{q_{i}r_{j}}\partial_{q_{j}}\varepsilon],
\end{align*}
where $\theta_{l}^{0}=\langle u_{n}|\hat{\theta}_{l}|u_{n}\rangle$. Since the
driving $\boldsymbol{F}$ field does not equilibrate with the electron system,
the argument of the equilibrium distribution function remains as $\varepsilon
$. Some manipulations similar to those leading to Eq. (\ref{M edge}) give rise
to $\boldsymbol{j}^{\text{e}}=\boldsymbol{j}^{\text{e,tr}}+\boldsymbol{\nabla
}\times\boldsymbol{M}^{\text{N}}$, with the transport electric current reading
\ \
\begin{equation}
j_{i}^{\text{e,tr}}=e\int f\left(  \varepsilon\right)  \left[  \partial
_{q_{i}}d_{jl}^{\theta}+\Omega_{q_{i}q_{j}}\theta_{l}^{0}\right]  \partial
_{j}F_{l}. \label{je general}%
\end{equation}
In the case of $\boldsymbol{\hat{\theta}}=e$ and $\boldsymbol{F}=\phi$,
$\boldsymbol{j}^{\text{e,tr}}$ reduces to the intrinsic anomalous Hall current
$j_{i}^{\text{e,tr}}=-e^{2}\int f\left(  \varepsilon\right)  \Omega
_{q_{i}q_{j}}E_{j}$. When $\boldsymbol{\hat{\theta}}$ is not a conserved
quantity, its dipole moment contributes to the transport current driven by the
$\boldsymbol{F}$-field gradient.

After some similar manipulations, the local thermal current density up to the
first order of spatial gradients is obtained as
\begin{align}
\boldsymbol{j}^{\text{h}}  &  =\int s\left(  \varepsilon\right)  T\left[
\partial_{\boldsymbol{q}}\delta\varepsilon+\theta_{i}^{0}\boldsymbol{\nabla
}F_{i}\times\boldsymbol{\Omega}_{\boldsymbol{q}}\right] \nonumber\\
&  +\boldsymbol{\nabla}\times\left[  \boldsymbol{M}^{\text{h}}+\int\theta
_{i}^{0}F_{i}(f\boldsymbol{m}^{\text{N}}+g\boldsymbol{\Omega}_{\boldsymbol{q}%
})\right]  ,
\end{align}
where the second line is the thermal magnetization current density in the
presence of the perturbation $\boldsymbol{\hat{\theta}}\cdot\boldsymbol{F}$.
In the case that $\boldsymbol{\hat{\theta}}=e$ and $\boldsymbol{F}=\phi$, the
above $\boldsymbol{j}^{\text{h}}$ reduces to Eq. (\ref{jh-local}) in the
absence of statistical forces. The transport thermal current is then
identified as the first line of the above equation, namely%
\begin{equation}
j_{i}^{\text{h,tr}}=\int s\left(  \varepsilon\right)  T\left[  \partial
_{q_{i}}d_{jl}^{\theta}+\Omega_{q_{i}q_{j}}\theta_{l}^{0}\right]  \partial
_{j}F_{l}.
\end{equation}
As the entropy density $s\left(  \varepsilon\right)  $\ goes to zero at the
zero-temperature limit, the above transport thermal current behaves well in
this limit.

One could then ask if there is the Mott relation linking the transport thermal
and electric currents when they are driven by the first-order spatial gradient
of the $\boldsymbol{F}$ field. For these two kinds of current, textbooks and
previous studies only proved the Mott relation in the case that an electric
field is the driving force. Here the present theory enables us to extend the
regime of validity of the Mott relation to the case where the $\boldsymbol{F}%
$-field gradient serves as the driving force. In fact, when the temperature is
much less than the distances between the chemical potential and band edges,
the Sommerfeld expansion can be used \cite{Xiao2016}, yielding the entropy
density $s\left(  \varepsilon\right)  =\frac{1}{3}\pi^{2}k_{B}^{2}%
T\delta\left(  \mu-\varepsilon\right)  $. Then we arrive at the Mott relation
\begin{equation}
\frac{\boldsymbol{j}^{\text{h,tr}}}{T}=\frac{\pi^{2}k_{B}^{2}T}{3e}%
\frac{\partial\boldsymbol{j}^{\text{e,tr}}\left(  \epsilon\right)  }%
{\partial\epsilon}|_{\epsilon=\mu},
\end{equation}
where $\boldsymbol{j}^{\text{e,tr}}\left(  \epsilon\right)  $ is the
zero-temperature transport electric current with Fermi energy $\epsilon$.

\section{Conclusion and Discussion}

In conclusion, we have provided a semiclassical description for the linear
thermal transport induced by the electric field and the gradients of chemical
potential and temperature in the presence of material inhomogeneity, based on
pure bulk considerations. In our method, applying simply the semiclassical
equations of motion in the presence of material inhomogeneity leads to a
systematic and efficient approach to both the magnetization and transport
currents. The results that previously can only be obtained in different
theories even in the absence of material inhomogeneity now emerge in a unified
theory. As long as the Berry-curvature-modified semiclassical equations of
motion hold for the considered quasi-particles, the linear intrinsic thermal
transport falls into the present framework, irrespective of the specific
content of Berry curvatures in different physical contexts
\cite{Xiao2006,Zhang2008,Cheng2016,Murakami2011PRL,Zhang2016,Zhang2019,Park2019}%
.

By introducing a fictitious inhomogeneity, this theory also applies to
homogeneous samples. Therefore, our theory promotes the material inhomogeneity
to be a basic and generic trick in the unified semiclassical theory of the
intrinsic linear thermoelectric and thermal transport. The trick of
inhomogeneity is independent of the driving force of transport, thus we have
included more generic mechanical driving forces other than the electric field,
and established the Mott relation between the resultant transport thermal and
electric currents.

Our theory can be extended further in several interesting directions. An
example is the thermal Hall effect mediated by Bogoliubov quasi-particles,
i.e., bogolons, in superconductors with time-reversal broken pairing like
$d+id$ \cite{Zhang2008}. As a momentum-dependent composition of electrons and
holes, the bogolon does not possess a definite charge, but has a definite
spin. Thus the scalar potential perturbation in the electronic Hamiltonian is
endowed with a spin structure in Nambu space, where the bogolons live in
\cite{Liang2017}. Therefore, the usual scalar potential felt by electrons in
superconductors is no longer a scalar one for bogolons. As a consequence, the
application of the boundary-current theory, where the quasi-particle
wave-packet is assumed to feel the gradient of the scalar potential provided
by the boundary, becomes very subtle for bogolon wave-packets. On the other
hand, in our theory the spin structure of the fictitious $\boldsymbol{w}$
field, which can be a scalar, vector or tensor field, is arbitrary. Thereby,
to apply our theory to the intrinsic thermal transport of bogolons is
appealing. To do this, one should derive the equations of motion
(\ref{EOM}) for bogolons, which has not been done up to now. Since this
subject would involve much more derivations and notations arising from the
characteristics of superconductors, it deserves a separate elaborate study.

Another issue of current interest is the electric and thermal current
responses to the gradient of an external electric field
\cite{Gao2019,Lapa2019}, where the orbital and thermal magnetization currents,
under the perturbation of the $\boldsymbol{E}$-field gradient, should be
discounted to acquire the transport currents. While the orbital magnetization
current can be accounted for along the approach of Ref. \cite{Xiao2006}, as
was done recently \cite{Gao2019}, the thermal magnetization current cannot. We
can show that our theory by using a fictitious inhomogeneity can be
generalized to obtain both the thermal and orbital magnetization current in
this case, and the resulting transport thermal current at the order of
$\boldsymbol{\nabla E}$ still obeys the Mott relation with the transport
electric current. However, such a generalization needs much more theoretical
input from the nonlinear semiclassical theory ($\boldsymbol{\nabla E}$ is the
second-order gradient of the electrostatic potential), which goes beyond the scope of
the present linear-response one and thus will be presented in a separate work.

In addition, the nonlinear generalization of our theory also allows for
attaining the linear thermoelectric responses of the orbital magnetization in
two-dimensional metallic systems with reduced symmetry \cite{Xiao2020}. This
potential for thermoelectric generation and control of magnetization via the
orbital degree of freedom would be of current great interest in low-symmetry
two-dimensional antiferromagnets.

\begin{acknowledgments}
We thank Liang Dong, Bangguo Xiong, Yinhan Zhang and Yang Gao for insightful discussions.
Q.N. is supported by DOE (DE-FG03-02ER45958, Division of Materials
Science and Engineering) on the geometric formulation in this work.
\end{acknowledgments}


\begin{thebibliography}{99}                                                                                               %


\bibitem {Xiao2006}D. Xiao, Y. Yao, Z. Fang, and Q. Niu, Phys. Rev. Lett.
\textbf{97}, 026603 (2006).

\bibitem {Zhang2008}C. Zhang, S. Tewari, V. M. Yakovenko, and S. Das Sarma,
Phys. Rev. B \textbf{78}, 174508 (2008).

\bibitem {Jauho2015}X.-Q. Yu, Z.-G. Zhu, G. Su, and A.-P. Jauho, Phys. Rev.
Lett. \textbf{115}, 246601 (2015).

\bibitem {Cheng2016}R. Cheng, S. Okamoto, and D. Xiao, Phys. Rev. Lett.
\textbf{117}, 217202 (2016).

\bibitem {Zhang2010}L. Zhang, J. Ren, J.-S. Wang, and B. Li, Phys. Rev. Lett.
\textbf{105}, 225901 (2010).

\bibitem {Murakami2011PRL}R. Matsumoto and S. Murakami, Phys. Rev. Lett.
\textbf{106}, 197202 (2011).

\bibitem {Murakami2011PRB}R. Matsumoto and S. Murakami, Phys. Rev. B
\textbf{84}, 184406 (2011).

\bibitem {Qin2012}T. Qin, J. Zhou, and J. Shi, Phys. Rev. B \textbf{86},
104305 (2012).

\bibitem {Murakami2014}R. Matsumoto, R. Shindou, and S. Murakami, Phys. Rev. B
\textbf{89}, 054420 (2014).

\bibitem {Lee2015}H. Lee, J. H. Han, and P. A. Lee, Phys. Rev. B \textbf{91},
125413 (2015).

\bibitem {Zhang2016}L. Zhang, New J. Phys. \textbf{18}, 103039 (2016).

\bibitem {Saito2019}T. Saito, K. Misaki, H. Ishizuka, and N. Nagaosa, Phys.
Rev. Lett. \textbf{123}, 255901 (2019).

\bibitem {Zhang2019}X. Zhang, Y. Zhang, S. Okamoto, and D. Xiao, Phys. Rev.
Lett. \textbf{123}, 167202 (2019).

\bibitem {Park2019}S. Park and B.-J. Yang, Phys. Rev. B \textbf{99}, 174435 (2019).

\bibitem {Go2019}G. Go, S. K. Kim, and K.-J. Lee, Phys. Rev. Lett.
\textbf{123}, 237207 (2019).

\bibitem {Cooper1997}N. R. Cooper, B. I. Halperin, and I. M. Ruzin, Phys. Rev.
B \textbf{55}, 2344 (1997).

\bibitem {Luttinger1964}J. M. Luttinger, Phys. Rev. \textbf{135}, A1505 (1964).

\bibitem {Streda1977}L. Smr\v{c}ka and P. St\v{r}eda, J. Phys. C \textbf{10},
2153 (1977).

\bibitem {Qin2011}T. Qin, Q. Niu, and J. Shi, Phys. Rev. Lett. \textbf{107},
236601 (2011).

\bibitem {Tatara2015}G. Tatara, Phys. Rev. Lett. \textbf{114}, 196601 (2015).

\bibitem {Shitade2014}A. Shitade, Prog. Theor. Exp. Phys. \textbf{2014},
123I01 (2014).

\bibitem {Xiao2009}D. Xiao, J. Shi, D. P. Clougherty, and Q. Niu, Phys. Rev.
Lett. \textbf{102}, 087602 (2009).

\bibitem {Sundaram1999}G. Sundaram and Q. Niu, Phys. Rev. B \textbf{59}, 14915 (1999).

\bibitem {Xiao2010}D. Xiao, M.-C. Chang, and Q. Niu, Rev. Mod. Phys.
\textbf{82}, 1959 (2010).

\bibitem {Shi2007}J. Shi, G. Vignale, D. Xiao, and Q. Niu, Phys. Rev. Lett.
\textbf{99}, 197202 (2007).

\bibitem {Ziman1972}J. M. Ziman, {\itshape}Principles of the Theory of Solids
(Cambridge University Press, Cambridge, 1972).

\bibitem {Ashcroft}N. W. Ashcroft and N. D. Mermin, Solid State Physics
(Saunders, Philadelphia, 1976).

\bibitem {Xiao2005}D. Xiao, J. Shi, and Q. Niu, Phys. Rev. Lett. \textbf{95},
137204 (2005).

\bibitem {Streda1983}P. St\v{r}eda and L. Smr\v{c}ka, J. Phys. C \textbf{16},
L895 (1983).

\bibitem {Nagaosa2012}K. Nomura, S. Ryu, A. Furusaki, and N. Nagaosa, Phys.
Rev. Lett. \textbf{108}, 026802 (2012).

\bibitem {Xiao2016}C. Xiao, D. Li, and Z. Ma, Phys. Rev. B \textbf{93}, 075150 (2016).

\bibitem {Liang2017}L. Liang, S. Peotta, A. Harju, and P. Torma, Phys. Rev. B
\textbf{96}, 064511 (2017).

\bibitem {Gao2019}Y. Gao and D. Xiao, Phys. Rev. Lett. \textbf{122}, 227402 (2019).

\bibitem {Lapa2019}M. F. Lapa and T. L. Hughes, Phys. Rev. B \textbf{99},
121111(R) (2019).

\bibitem {Xiao2020}C. Xiao and Q. Niu, arXiv:2002.01637
\end{thebibliography}
\end{document}